# Microwave properties of $Nd_{0.5}Sr_{0.5}MnO_3$: a key role of the $(x^2-y^2)$-orbital effects

S. Zvyagin[1], A. Angerhofer[1], K.V. Kamenev[2], L.-C. Brunel[3], G. Balakrishnan[4], and D.M$^c$K.Paul[4]

[1]Department of Chemistry, The University of Florida, PO Box 117200, Gainesville, FL 32611-7200, USA
[2]Department of Physics and Astronomy, The University of Edinburgh, Mayfield Road, Edinburgh EH9 3JZ, UK
[3]National High Magnetic Field Laboratory, Florida State University, 1800 E. Paul Dirac Dr., Tallahassee, FL32310, USA
[4]Department of Physics, University of Warwick, Coventry CV4 7AL, UK

Transmittance of the colossal magnetoresistive compound $Nd_{0.5}Sr_{0.5}MnO_3$ showing metal-insulator phase transition has been studied by means of the submm- and mm-wavelength band spectroscopy. An unusually high transparency of the material provided direct evidence for the significant suppression of the coherent Drude weight in the ferromagnetic metallic state. Melting of the A-type antiferromagnetic states has been found to be responsible for a considerable increase in the microwave transmission, which was observed at the transition from the insulating to the metallic phase induced by magnetic field or temperature. This investigation confirmed a dominant role of the $(x^2-y^2)$-orbital degree of freedom in the low-energy optical properties of $Nd_{0.5}Sr_{0.5}MnO_3$ and other doped manganites with planar $(x^2-y^2)$-orbital order, as predicted theoretically. The results are discussed in terms of the orbital-liquid concept.

Hole-doped manganese oxides with distorted perovskite structure $R_{1-x}D_xMnO_3$ (where R and D are trivalent rare earth and divalent alkaline earth ions, respectively) have attracted significant interest due to their unique electronic and magnetic properties. It is known, that orbital effects play an important role in the physics of these compounds (for a review see, for instance, reference [1]), and may determine the main peculiarities of the electron-transfer interactions in manganites. One of the most intriguing members of this family of materials is $Nd_{0.5}Sr_{0.5}MnO_3$. A number of fascinating phenomena have been recently discovered and studied in $Nd_{0.5}Sr_{0.5}MnO_3$ (see, for example [2-4]), that made this compound extremely attractive from the point of view of testing recent theoretical predictions, including those involving charge and orbital effects.

Optical reflectivity studies play a central role in the determination of the electronic and orbital states of rare-earth manganese oxides. Far-infrared (FIR) spectroscopy plays an especial role, providing important information about low-energy optical properties and characteristics of the manganites, including the Drude-weight contribution and, thus, the metallicity of these materials. Unfortunately, reflectivity spectra are very sensitive to the quality of the sample surface, that eventually may cause sometimes rather debatable and confused results (see, for instance, the discussion in reference [5]). The primary motivation for our study was using transmission-probe mm- and submm- wavelength band spectroscopy techniques to investigate the bulk low-energy optical properties of $Nd_{0.5}Sr_{0.5}MnO_3$ at the metal-insulator phase transition, and link them to its magnetic properties.

Let us first briefly describe magnetic properties of $Nd_{0.5}Sr_{0.5}MnO_3$. The compound undergoes a series of magnetic phase transitions on cooling. At ambient temperature $Nd_{0.5}Sr_{0.5}MnO_3$ is paramagnetic (PM) and insulating. At a temperature of about 250 K ($T_C$) it becomes a ferromagnetically (FM) ordered metal. At yet lower temperatures of about 160 K the sample undergoes a transition into a charge-ordered (CO) insulating state [4], characterized by the co-existence of A-type antiferromagnetic (AF-A) and CE-type antiferromagnetic (AF-CE) phases [3,6]. The existence of the AF-A phase in samples with nominal composition $Nd_{0.5}Sr_{0.5}MnO_3$ can be attributed to a slight excess of $Mn^{4+}$ [7], that indicates an extremely delicate competition between FM, AF-CE and AF-A states in $Nd_{0.5}Sr_{0.5}MnO_3$ with x~0.5. The development of the AF-A phase starts at a temperature, higher than the temperature of the transition into the AF-CE phase [3,6]. The key feature of the AF-CE phase is the alternate ordering of the $Mn^{3+}$ and $Mn^{4+}$ ions into a CO state, which is accompanied by the $(3z^2-r^2)$-orbital ordering (OO) [2]. In the AF-A phase the spins are ordered ferromagnetically in the *ab* plane of an orthorhombically distorted pseudo-cubic structure (symmetry group *Ibmm*). The localized moments on the Mn sites point along the *a* axis, and these ferromagnetic planes are stacked antiferromagnetically along the *c* axis. The AF-A phase is accompanied by in-plane $(x^2-y^2)$-orbital ordering [8]. This in-plane ordering is so strong that even in the fully spin-polarized ferromagnetic (FM) state quasi-two-dimensional (2D) spin fluctuations still exist, reflecting the underlying planar $(x^2-y^2)$-type orbital correlations [6,9]. The AF-A states in doped manganites are characterized by a pronounced metallicity of the *ab* planes [10]. The motion of the charge carriers is quasi-2D since there is no hopping along the *c* axis.

In our experiments we used single-crystals of $Nd_{0.5}Sr_{0.5}MnO_3$, which have been grown and characterized as described elsewhere [12]. The normalized microwave transmittance of $Nd_{0.5}Sr_{0.5}MnO_3$ at a frequency of 288 GHz at different temperatures is shown as a function of magnetic field in Fig.1. At all three temperatures (T = 120, 86 and 40 K) and in zero field the sample is in the antiferromagnetic phase. The sample undergoes a transition to the metallic FM state in a high magnetic field [4]. Two remarkable features were observed. First, we found that the samples with a relatively large thickness (of about 5 mm) were quite transparent for radiation in the FM metallic state. Note, that the values for resistivity of $Nd_{0.5}Sr_{0.5}MnO_3$ in in this phase are relatively close to the numbers characteristic of regular metals (~$10^{-4}$ Ohm cm, [4]). We estimate that the power attenuation by the sample in the FM metallic phase was about 37 dB/cm. This value is typical for dielectrics, rather than for ordinary metals with a well-defined skin-depth of the order of micrometers in this spectral range. Observation of a strong ferromagnetic resonance line in the FM phase (as it can be seen, for instance, in the Fig.1 at the field of about 10.3 T) illustrates unusually high transparency of the sample in the microwave region as well. Our second intriguing observation was that the sample was more transparent for microwaves in the high-field FM than in the low-field AF phase. The double-exchange theory predicts that the closer the alignment of the spins to parallel on the adjacent Mn atoms, the easier it is for the hopping of a charge carrier between the Mn sites [12]. In other words, the material in the FM phase should be a better conductor (and thus have a stronger electron contribution to microwave absorption) than when it is in the PM and AF phases. However, we observe lower transparency for the low-field AF state, which suggests that the AF phase is screening the microwaves.

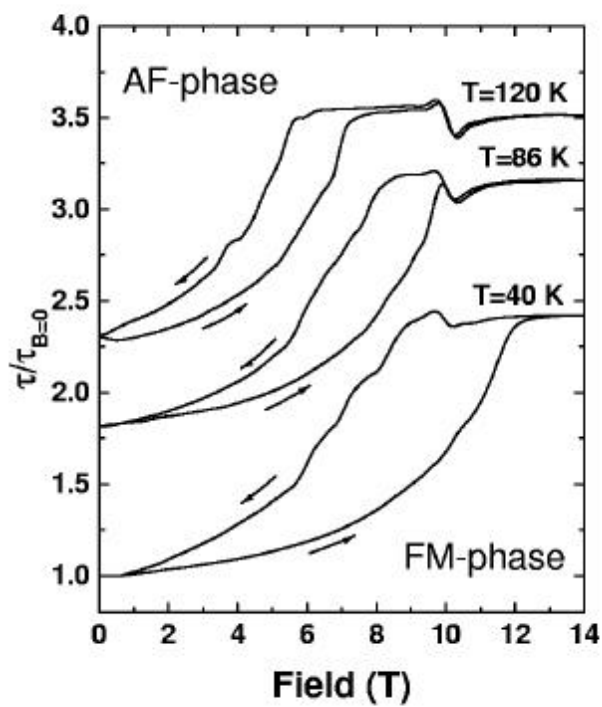

Fig. 1. Normalized microwave transmittance of $Nd_{0.5}Sr_{0.5}MnO_3$ at a frequency of 288 GHz and temperatures T = 40, 86, and 120 K (the arrows indicate the increasing and decreasing field runs; spectra are offset for clarity).

The measurements were repeated at different frequencies in the frequency range of 90-500 GHz and a similar magnetic field behavior was observed. The results of these studies are summarized in Fig.2, where the normalized microwave transmittance at T = 120 K is plotted as a function of magnetic field at different frequencies. We observed that the change in the transparency is dependent on the radiation frequency - lower frequency induces a more pronounced difference in the transparency (Fig. 2). The microwave transmittance clearly shows a field-dependent hysteresis, which is also dependent on temperature (Fig. 1). Again, the change in the microwave transmission detected below $T_{CO}$ ~ 160 K in an applied magnetic field corresponds to the collapse of the AF state, followed by a transition to the high-field FM phase. One can see that the transmitted signal is saturated in high magnetic fields. The saturation fields $B_{c1}$, $B_{c2}$ (Fig. 2) were chosen to mark the field-temperature phase diagram, shown in Fig. 2, inset. The values of the critical fields determined from the microwave investigation are in a good agreement with the data displaying the obtained earlier [4] electronic phase diagram for the metal-insulator transitions.

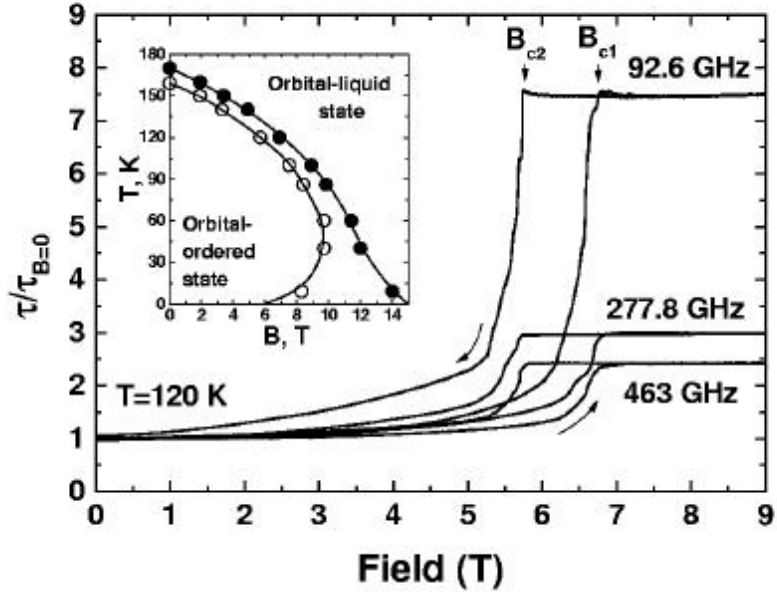

Fig. 2. Normalized microwave transmittance at a temperature of T = 120 K and various frequencies (the arrows indicate the increasing and decreasing field runs). The inset shows the temperature-field phase diagram of $Nd_{0.5}Sr_{0.5}MnO_3$, obtained from the microwave measurements. The solid and open circles indicate data obtained in ascending and descending magnetic fields respectively. The lines are guides for the eye.

The zero-field temperature dependence also shows a change in the microwave transmission. The normalized transmission at the frequency of 92.6 GHz as a function of temperature is shown in Fig. 3. At ambient temperature the sample has a high transmittance, which correlates well with the fact that the sample is in the insulating PM state. On the transition to the FM phase at $T_C$ ~ 250 K the microwave transmittance starts decreasing, reflecting the metallicity of the FM phase. At a temperature of about 210 K we observe a break in the slope of the intensity vs temperature curve, which we ascribe to the onset of the AF-A phase [6]. This temperature is denoted as $T_N^A$ in Fig. 3. In the temperature region between 170 K and 160 K the transmittance decreases rapidly, indicating the collapse of the FM state and the transition to a fully developed AF state.

The microwave transmittance then remains almost unchanged in the broad temperature region of 140 – 5 K (Fig. 3). The dependence of the microwave transmittance on temperature reflects the range of existence and temperature behavior of the AF-A phase, previously studied by neutron scattering measurements. From this we conclude that the AF-A phase, accompanied by $(x^2-y^2)$-type OO, could be responsible for the anomalous behaviour observed in $Nd_{0.5}Sr_{0.5}MnO_3$.

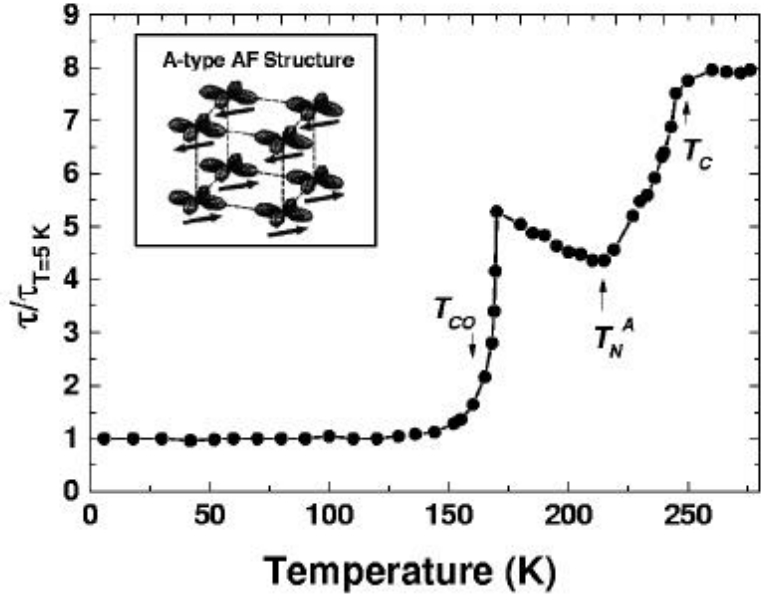

Fig. 3. Normalized microwave transmittance of $Nd_{0.5}Sr_{0.5}MnO_3$ as a function of increasing temperature at a frequency of 92.6 GHz. The lines are guides for the eye. The inset shows a schematic model of the ordering of the $e_g$-orbitals in the AF-A phase.

Summarizing our experimental findings, we would like to remind the reader of the two main discoveries we have made: (i) the transparency of the FM phase to the microwave radiation (even though it is a metallic phase), and (ii) higher transparency in the FM phase than the AF phase.

To explain our observations we turn to the recent theoretical studies of optical conductivity of quasi 2D FM states in the AF-A spin structure. These states are due to the strong magnetic coupling of Mn ions in each $ab$ plane and a weaker coupling between the adjacent planes along the $c$ axis. The AF-A spin structure is accompanied by the planar $(x^2-y^2)$-type OO (as depicted in Fig. 3 (inset)), which can be significant for the magnetic and electron-transport properties of manganites [1]. As we mentioned above, the planar-type $(x^2-y^2)$-orbital ordering may have a crucial effect on the dynamic spin correlations in $Nd_{0.5}Sr_{0.5}MnO_3$. Even in the fully spin-polarized FM state and the high-temperature PM state the $(x^2-y^2)$-type orbital degree of freedom may play an important role causing quasi-2D orbital fluctuations and establishing the two-dimensionality of the spin dynamics [6,9]. Some important features of the quasi-2D orbital fluctuations in the FM state and their effect on the optical conductivity (including the low-energy range) have been studied theoretically [13-15]. Using a numerical approach Ishihara *et al.* [13] concluded that the quasi-2D nature of the orbital fluctuations in a 3D cubic system leads to a so-called orbital liquid state. Although the amplitude of the short-range order in this state can be large, the fluctuations of the orbital degrees of freedom are large enough to prevent the OO. A microscopic approach to the problem, which combines strong correlations and orbital degeneracy, has been performed by Kilian [14]. The theories consider the role of the orbital degree of freedom coupled to the lattice at the metal-insulator transition, which is supposed to determine the main features of the optical conductivity in manganites. It was shown, that the *dynamic orbital fluctuations could significantly diffuse the charge carrier motion, strongly reducing the coherent component of the optical conductivity* (normally seen as a Drude peak in the optical spectra) and transferring the spectral weight to the high-frequency region. Thus scattering from the orbital fluctuations causes the change in the low-energy optical properties of $Nd_{0.5}Sr_{0.5}MnO_3$ at the insulator-to-metal phase transition (that manifests itself in the microwave region), making samples more transparent to radiation in the FM state.

In contrast, the metallicity of AF-A phase can be expected in the low-energy optical spectral region, contributing to microwave attenuation. It was theoretically predicted [16] that due to its metallicity $(x^2-y^2)$-orbital ordered regime should exhibit a non-zero Drude-like component. Sudden and relatively large increase of the microwave attenuation in the vicinity of the metal-to-insulator phase transition (that is, in fact, a transition from 3D to 2D metal) confirmed this suggestion. Observation of critical points in temperature which coincide with the temperature behavior of AF-F phase supports this prediction as well.

Thus, the microwave behavior of $Nd_{0.5}Sr_{0.5}MnO_3$ can be explained in terms of the following scenario. At low temperatures (T < $T_{CO}$, Fig.3) the $(x^2-y^2)$-orbital ordered 2D-metallic AF-A phase dominates the optical conductivity, giving rise to the low-energy electronic conductivity within the $(x^2-y^2)$-orbital ordered planes, and also determines the low-temperature microwave absorption. The 2D orbital fluctuations result in the orbital liquid behavior of the FM phase, suppressing the Drude peak, and, thus, causing a pronounced increase in the microwave transmission at T > $T_{CO}$. The decrease in the microwave transmission in the temperature range of ~ 170K – 210 K reflects the corresponding weakening of the AF-A correlations [3,6], which makes the sample more "metallic" and less transparent to the microwaves. At temperatures above T ~ 220 K thermal fluctuations start to suppress the FM states, increasing the PM contribution and making the sample most transparent at T > $T_C$. Application of an external magnetic field below $T_{CO}$ also leads to the melting of the 2D-metallic AF-A states. Similarly, the $(x^2-y^2)$-orbital fluctuations diffuse the motion of the charge carriers in the high-field FM phase, providing better conditions for the propagation of microwaves (Fig. 1, Fig. 2).

Similar microwave behavior has been recently observed in $Pr_{0.5}Sr_{0.5}MnO_3$ [17], which is also characterized by the A-type AFM structure. An unusual (nonmetallic) character of the FM phase has been observed in some manganites using FIR methods. In particular, a strong suppression of the Drude weight in the FM phase was reported [18-

20]. No clear evidence of the coherent component was observed in the high-temperature FM phase of $Nd_{0.5}Sr_{0.5}MnO_3$ [21]. These findings are in a good agreement with our results supporting the significant role of $(x^2-y^2)$-orbital effects.

In conclusion, we have reported on studies of the bulk properties of a single-crystals of $Nd_{0.5}Sr_{0.5}MnO_3$ in a wide microwave spectral range and at a variety of magnetic fields and temperatures. We have for the first time observed anomalous microwave transmission in the ferromagnetic state of this manganese oxide. A considerable increase of the transmittance has been observed at the transition from the insulating to the metallic phase induced by magnetic field or temperature. The explanation for these effects has been given by taking into account some characteristic features of the AF-A structure and the planar $(x^2-y^2)$-orbital ordering. Our observations strongly support the orbital-liquid concept in perovskite transition-metal oxides. The results of the present study also reveal the potential of $Nd_{0.5}Sr_{0.5}MnO_3$ and related compounds for the construction of field/temperature-controlled high-frequency switches, modulators and other elements used for microwave techniques.

S.Zvyagin is grateful for the support from Alexander von Humboldt Foundation and very thankful to Prof. B.Lüthi for hospitality. The British co-authors acknowledge the support through the EPSRC Grants No.GR/K95802 and No.GR/M75471.